# From "Identical" to "Similar": Fusing Retrieved Lists Based on Inter-Document Similarities

**Anna Khudyak Kozorovitsky**                     ANNAK@TX.TECHNION.AC.IL
**Oren Kurland**                                  KURLAND@IE.TECHNION.AC.IL
*Faculty of Industrial Engineering and Management*
*Technion — Israel Institute of Technology*

## Abstract

Methods for *fusing* document lists that were retrieved in response to a query often utilize the retrieval scores and/or ranks of documents in the lists. We present a novel fusion approach that is based on using, in addition, information induced from *inter-document similarities*. Specifically, our methods let similar documents from different lists provide relevance-status support to each other. We use a graph-based method to model relevance-status propagation between documents. The propagation is governed by inter-document-similarities and by retrieval scores of documents in the lists. Empirical evaluation demonstrates the effectiveness of our methods in fusing TREC *runs*. The performance of our most effective methods transcends that of effective fusion methods that utilize only retrieval scores or ranks.

## 1. Introduction

The ad hoc retrieval task is to find the documents most pertaining to an information need underlying a given query. Naturally, there is considerable uncertainty in the retrieval process — e.g., accurately inferring what the "actual" information need is. Thus, researchers proposed to utilize different information sources and types to address the retrieval task (Croft, 2000b). For example, utilizing multiple document representations (Katzer, McGill, Tessier, Frakes, & Dasgupta, 1982), query representations (Saracevic & Kantor, 1988; Belkin, Cool, Croft, & Callan, 1993), and search techniques (Croft & Thompson, 1984; Fox & Shaw, 1994), have been proposed as a means for improving retrieval effectiveness.

Many of the approaches just mentioned depend on the ability to effectively *fuse* several retrieved lists so as to produce a single list of results. Fusion might be performed under a single retrieval system (Croft & Thompson, 1984), or upon the results produced by different search systems (Fox & Shaw, 1994; Callan, Lu, & Croft, 1995; Dwork, Kumar, Naor, & Sivakumar, 2001). Conceptually, fusion can be viewed as integrating "*experts' recommendations*" (Croft, 2000b), where the expert is a retrieval model used to produce a ranked list of results — the expert's recommendation.

A principle underlying many fusion methods is that the documents that are highly ranked in many of the lists, i.e., that are highly "recommended" by many of the "experts", should be ranked high in the final result list (Fox & Shaw, 1994; Lee, 1997). The effectiveness of approaches that employ this principle often depends on the overlap[1] between non-relevant documents in the lists being much smaller than that between relevant documents (Lee, 1997). However, several studies have shown that this is often not the case, more specifically, that on many occasions there are (many) different relevant documents across the lists to be fused (Das-Gupta & Katzer, 1983; Griffiths, Luckhurst, & Willett, 1986; Chowdhury, Frieder, Grossman, & McCabe, 2001; Soboroff, Nicholas, & Cahan, 2001; Beitzel et al., 2003).

---

1. We use the term "overlap" to refer to the number of documents shared by the retrieved lists rather than in reference to content overlap.

We propose a novel approach to fusion of retrieved lists that addresses, among others, the relevant-documents mismatch issue just mentioned. A principle guiding the development of our methods is that *similar documents* — from different lists, as well as those in the same list — can provide relevance-status support to each other, as they potentially discuss the same topics (Shou & Sanderson, 2002; Baliński & Daniłowicz, 2005; Diaz, 2005; Kurland & Lee, 2005; Meister, Kurland, & Kalmanovich, 2010). Specifically, if relevant documents are assumed to be similar following the *cluster hypothesis* (van Rijsbergen, 1979), then they can provide "support" to each other via inter-document similarities.

Inspired by work on re-ranking a *single* retrieved list using inter-document similarities within the list (Baliński & Daniłowicz, 2005; Diaz, 2005; Kurland & Lee, 2005), our approach uses a graph-based method to model relevance-status propagation between documents in the lists to be fused. The propagation is governed by inter-document-similarities and by the retrieval scores of documents in the lists. Specifically, documents that are highly ranked in the lists, and are similar to other documents that are highly ranked, are rewarded. If inter-document-similarities are not utilized — i.e., only retrieval scores are used — then some of our methods reduce to standard fusion approaches.

Empirical evaluation shows that our methods are highly effective in fusing TREC *runs* (Voorhees & Harman, 2005); that is, lists of document that were created in response to queries by search systems that participated in TREC. Our most effective methods post performance that is superior to that of effective standard fusion methods that utilize only retrieval scores. We show that these findings hold whether the runs to be fused, which are selected from all available runs per track (challenge) in TREC, are the most effective ones, or are randomly selected. Using an additional array of experiments we study the effect of various factors on the performance of our approach.

## 2. Fusion Framework

Let $q$ and $d$ denote a query and a document, respectively. We assume that documents are assigned with unique IDs; we write $d_1 \equiv d_2$ if $d_1$ and $d_2$ have the same ID, i.e., they are the same document. We assume that the document lists $L_1^{[q;k]}, \ldots, L_m^{[q;k]}$, or $L_1, \ldots, L_m$ in short, were retrieved in response to $q$ by $m$ retrievals performed over a given corpus, respectively; each list contains $k$ documents. We write $d \in L_i$ to indicate that $d$ is a member of $L_i$, and use $S_{L_i}(d)$ to denote the (positive) retrieval score of $d$ in $L_i$; if $d \notin L_i$ then $S_{L_i}(d) \stackrel{def}{=} 0$. The *document instance* $L_i^j$ is the document at rank $j$ in list $L_i$. To simplify notation, we often use $S(L_i^j)$ to denote the retrieval score of $L_i^j$ (i.e., $S(L_i^j) \stackrel{def}{=} S_{L_i}(L_i^j)$). The methods that we present consider the similarity $sim(d_1, d_2)$ between documents $d_1$ and $d_2$. The methods are not committed to a specific way of computing inter-document similarities. For example, the cosine measure between vector-space representations of documents can be used as in some previous work on re-ranking a single retrieved list (Diaz, 2005; Kurland & Lee, 2005). In Section 4.1 we describe our specific choice of a language-model-based inter-document similarity measure used for experiments following previous recommendations (Kurland & Lee, 2010).

## 2.1 Fusion Essentials

Our goal is to produce a single list of results from the retrieved lists $L_1, \ldots, L_m$. To that end, we opt to detect those documents that are "highly recommended" by the lists $L_1, \ldots, L_m$, or in other words, that are "*prestigious*" with respect to the lists. Given the virtue by which the lists were created, that is, in response to the query, we hypothesize that prestige implies relevance. The key challenge is then to formally define and quantify prestige.

Many current fusion approaches (implicitly) regard a document as prestigious if it is highly ranked in many of the lists. The CombSUM method (Fox & Shaw, 1994), for example, quantifies this prestige notion by summing the document retrieval scores[2]:

$$P_{CombSUM}(d) \stackrel{def}{=} \sum_{L_i : d \in L_i} S_{L_i}(d).$$

To emphasize even more the importance of occurrence in many lists, the CombMNZ method (Fox & Shaw, 1994; Lee, 1997), which is a highly effective fusion approach (Montague & Aslam, 2002), scales CombSUM's score by the number of lists a document is a member of:

$$P_{CombMNZ}(d) \stackrel{def}{=} \#\{L_i : d \in L_i\} \sum_{L_i : d \in L_i} S_{L_i}(d).$$

A potentially helpful source of information not utilized by standard fusion methods is *inter-document relationships*. For example, documents that are similar to each other can provide support for prestige as they potentially discuss the same topics. Indeed, work on re-ranking a single retrieved list has shown that prestige induced from inter-document similarities is connected with relevance (Kurland & Lee, 2005). In the multiple-lists setting that we address here, information induced from inter-document similarities across lists could be a rich source of helpful information as well. A case in point, a document that is a member of a single list, but which is similar to other documents that are highly ranked in many of the lists could be deemed prestigious. Furthermore, similarity-based prestige can be viewed as a generalization of the prestige notion taken by standard fusion methods, if we consider documents to be similar if and only if they are the same document.

## 2.2 Similarity-Based Fusion

We use graphs to represent propagation of "prestige status" between documents; the propagation is based on inter-document similarities and retrieval scores. The nodes of a graph represent either documents, or document instances (appearances of documents) in the retrieved lists. In the latter case, the same document can be represented by several nodes, each of which corresponds to its appearance in a list, while in the former case, each node corresponds to a different document.

---

2. We assume that retrieval scores are normalized for inter-list compatibility; details about the normalization scheme we employ in the experiments are provided in Section 4.1.

The development of the following graph-construction method and prestige-induction technique is inspired by work on inducing prestige in a single retrieved list (Kurland & Lee, 2005). In contrast to this work, however, we would like to exploit the special characteristics of the fusion setup. That is, the fact that documents can appear in several retrieved lists with different retrieval scores that might be produced by different retrieval methods. Accordingly, all fusion methods that we develop are novel to this study.

Formally, given a set of documents (document instances) $V$, we construct a weighted (directed) complete graph $G \stackrel{def}{=} (V, V \times V, wt)$ with the edge-weight function $wt$:

$$wt(v_1 \rightarrow v_2) \quad \stackrel{def}{=} \quad \begin{cases} sim(v_1, v_2) & \text{if } v_2 \in Nbhd(v_1; \alpha), \\ 0 & \text{otherwise;} \end{cases}$$

$v_1, v_2 \in V$; and, $Nbhd(v; \alpha)$ is the $\alpha$ elements $v'$ in $V - \{v'' : v'' \equiv v\}$ that yield the highest $sim(v, v')$ — i.e., $v$'s nearest neighbors in $V$; $\alpha$ is a free parameter.[3] Previous work has demonstrated the merits of using directed nearest-neighbor-based graphs, as we use here, for modeling prestige-status propagation in setups wherein prestige implies "relevance to information need" (Kurland & Lee, 2005). (See Kurland, 2006 for elaborated discussion.)

As in work on inducing (i) journal prestige in bibliometrics (Pinski & Narin, 1976), (ii) Web-page prestige in Web retrieval (Brin & Page, 1998), and (iii) plain-text prestige for re-ranking a single list (Kurland & Lee, 2005), we can say that a node $v$ in $G$ is prestigious to the extent it receives prestige-status support from other prestigious nodes. We can quantify this prestige notion using $P(v; G) \stackrel{def}{=} \sum_{v' \in V} wt(v' \rightarrow v) P(v'; G)$. However, this recursive equation does not necessarily have a solution.

To address this issue, we define a smoothed version of the edge-weight function, which echoes PageRank's (Brin & Page, 1998) approach:

$$wt^{[\lambda]}(v_1 \rightarrow v_2) \stackrel{def}{=} \lambda \cdot \frac{\widehat{sim}(v_2, q)}{\sum_{v' \in V} \widehat{sim}(v', q)} + (1 - \lambda) \cdot \frac{wt(v_1 \rightarrow v_2)}{\sum_{v' \in V} wt(v_1 \rightarrow v')} \; ; \qquad (1)$$

$\lambda$ is a free parameter, and $\widehat{sim}(v, q)$ is $v$'s estimated query similarity. (Below we present various query-similarity measures.) The resultant graph is $G^{[\lambda]} \stackrel{def}{=} (V, V \times V, wt^{[\lambda]})$.

Note that each node in $G^{[\lambda]}$ receives prestige-status support to an extent partially controlled by the similarity of the document it represents to the query. Nodes that are among the nearest-neighbors of other nodes get an additional support. Moreover, $wt^{[\lambda]}$ can be thought of as a probability transition function, because the sum of weights on edges going out from a node is 1; furthermore, every node has outgoing edges to *all* nodes in the graph (self loops included). Hence, $G^{[\lambda]}$ represents an ergodic Markov chain for which a unique stationary distribution exists (Golub & Van Loan, 1996). This distribution, which can be found using, for example, the Power method (Golub & Van Loan, 1996), is the unique solution to the following prestige-induction equation under the constraint $\sum_{v' \in V} P(v'; G^{[\lambda]}) = 1$:

$$P(v; G^{[\lambda]}) \stackrel{def}{=} \sum_{v' \in V} wt^{[\lambda]}(v' \to v) P(v'; G^{[\lambda]}). \qquad (2)$$

---

3. Note that $Nbhd(v; \alpha)$ contains only nodes that represent documents that are not that represented by $v$.

| Algorithm | $V$ | $\widehat{sim}(v,q)$ | $Score(d)$ |
|---|---|---|---|
| SetUni | $\{d : d \in \bigcup_i L_i\}$ | $1$ | $P(d; G^{[\lambda]})$ |
| SetSum | $\{d : d \in \bigcup_i L_i\}$ | $P_{CombSUM}(v)$ | $P(d; G^{[\lambda]})$ |
| SetMNZ | $\{d : d \in \bigcup_i L_i\}$ | $P_{CombMNZ}(v)$ | $P(d; G^{[\lambda]})$ |
| BagUni | $\{L_i^j\}_{i,j}$ | $1$ | $\sum_{v \in V : v \equiv d} P(v; G^{[\lambda]})$ |
| BagSum | $\{L_i^j\}_{i,j}$ | $S(v)$ | $\sum_{v \in V : v \equiv d} P(v; G^{[\lambda]})$ |
| BagDupUni | $\{Dup(L_i^j)\}_{i,j}$ | $1$ | $\sum_{v \in V : v \equiv d} P(v; G^{[\lambda]})$ |
| BagDupMNZ | $\{Dup(L_i^j)\}_{i,j}$ | $S(v)$ | $\sum_{v \in V : v \equiv d} P(v; G^{[\lambda]})$ |

Table 1: Similarity-based fusion methods; $Score(d)$ is $d$'s final retrieval score.

| $L_1$ | $L_2$ |
|---|---|
| $L_1^1: d_1$ | $L_2^1: d_2$ |
| $L_1^2: d_2$ | $L_2^2: d_4$ |
| $L_1^3: d_3$ | $L_2^3: d_1$ |

Table 2: Example of two retrieved lists to be fused.

### 2.2.1 Methods

To derive specific fusion methods, we need to specify the graph $G^{[\lambda]}$ using which prestige is induced in Equation 2. More specifically, given the lists $L_1, \ldots, L_m$, we have to define a set of nodes $V$ that represent documents (or document instances); and, we have to devise a query-similarity estimate ($\widehat{sim}(v,q)$) to be used by the edge-weight function $wt^{[\lambda]}$ from Equation 1. The alternatives that we consider, which represent some ways of utilizing our graph-based approach, and the resultant fusion methods, are presented in Table 1. It is important to note that each fusion method produces a ranking of documents wherein a document cannot have more than one instance. To facilitate the discussion of the various methods from Table 1, we will refer to the example of fusing the two lists from Table 2, $L_1$ and $L_2$, each of which contains three documents.

The first group of methods does not consider occurrences of a document in multiple lists when utilizing inter-document similarities. Specifically, $V$, the set of nodes, is defined to be the <u>set</u>-union of the retrieved lists. For the example in Table 2, $V \stackrel{def}{=} \{d_1, d_2, d_3, d_4\}$. Thus, each document is represented in the graph by a single node. The prestige value of this node serves as the final retrieval score of the document. The **SetUni** method ignores the retrieval scores of documents by using a uniform query-similarity estimate; hence, only inter-document similarity information is utilized. The **SetSum** and **SetMNZ** methods, on the other hand, integrate also retrieval scores by using the CombSUM and CombMNZ prestige scores for query-similarity estimates, respectively.

The SetSum and SetMNZ methods are, in fact, generalized forms of CombSUM and CombMNZ, respectively. If we use the edge-weight function $wt^{[1]}$ (i.e., set $\lambda = 1$ in Equation 1), that is, do not exploit inter-document-similarity information, then SetSum and SetMNZ amount to CombSUM and CombMNZ, respectively; lower values of $\lambda$ result in more emphasis put on inter-document-similarities information. Furthermore, the set-based paradigm can be used so as to incorporate and generalize any fusion method by using the method's retrieval score as the query-similarity estimate. Then, setting $\lambda = 1$ amounts to using only the fusion method's retrieval scores. (See Appendix A for a proof.)

In contrast to the set-based methods, the bag-based methods consider occurrences of a document in multiple lists in utilizing inter-document similarity information. Each node in the graph represents an instance of a document in a list. Hence, the set of nodes ($V$) in the graph could be viewed as the bag-union of the retrieved lists. In the example from Table 2, $V \stackrel{def}{=} \{L_1^1, L_1^2, L_1^3, L_2^1, L_2^2, L_2^3\}$. The final retrieval score of a document is set to the sum of prestige scores of the nodes that represent it — i.e., that correspond to its instances in the lists. For example, the score of $d_1$ would be the sum of the scores of the nodes $L_1^1$ and $L_2^3$. It is also important to note that while the neighborhood set $Nbhd(v; \alpha)$ of node $v$ cannot contain nodes representing the same document represented by $v$, it can contain multiple instances of a different document. Thus, documents with many instances tend to receive more inter-document-similarity-based prestige-status support than documents with fewer instances.

The first representative of the bag-based methods, **BagUni**, ignores retrieval scores and considers only inter-document-similarities. Hence, BagUni differs from SetUni only by the virtue of rewarding documents with multiple instances. In addition to exploiting inter-document similarities, the **BagSum** method also uses the retrieval score of a document instance as the query-similarity estimate of the corresponding node. We note that Comb-SUM is a specific case of BagSum with $\lambda = 1$, as was the case for SetSum. (See Appendix A for a proof.) Furthermore, BagSum resembles SetSum in that it uses $\lambda$ for controlling the balance between using retrieval scores and utilizing inter-document similarities. However, documents with many instances get more prestige-status support in BagSum than in SetSum due to the bag-based representation of the lists.

Naturally, then, we opt to create a bag-based generalized version of the CombMNZ method. To that end, for *each* document instance $L_i^j$ that corresponds to document $d$, we define a new list $Dup(L_i^j)$. This list contains $n$ copies of $d$, each assigned to an arbitrary different rank between 1 and $n$ with $S(L_i^j)$ as a retrieval score; $n \stackrel{def}{=} \#\{L_i : d \in L_i\}$ — the number of original lists that $d$ belongs to. The set of nodes $V$ is composed of all document instances in the *newly* defined lists. For the example from Table 2, we get the following newly created lists:

| $L_a \stackrel{def}{=} Dup(L_1^1)$ | $L_b \stackrel{def}{=} Dup(L_2^3)$ | $L_c \stackrel{def}{=} Dup(L_1^2)$ | $L_d \stackrel{def}{=} Dup(L_2^1)$ | $L_e \stackrel{def}{=} Dup(L_1^3)$ | $L_f \stackrel{def}{=} Dup(L_2^2)$ |
|---|---|---|---|---|---|
| $L_a^1 : L_1^1 \equiv d_1$ | $L_b^1 : L_2^3 \equiv d_1$ | $L_c^1 : L_1^2 \equiv d_2$ | $L_d^1 : L_2^1 \equiv d_2$ | $L_e^1 : L_1^3 \equiv d_3$ | $L_f^1 : L_2^2 \equiv d_4$ |
| $L_a^2 : L_1^1 \equiv d_1$ | $L_b^2 : L_2^3 \equiv d_1$ | $L_c^2 : L_1^2 \equiv d_2$ | $L_d^2 : L_2^1 \equiv d_2$ | | |

The set of nodes, $V$, is $\{L_a^1, L_a^2, L_b^1, L_b^2, L_c^1, L_c^2, L_d^1, L_d^2, L_e^1, L_f^1\}$. Note, for example, that while $d_1$ was represented by a single node under the set-based representation, and by two nodes under the bag-based representation, here it is represented by four nodes. More generally, the number of nodes by which each document is represented here is the square of the number of appearances of the document in the lists.

The **BagDupUni** method, then, uses a uniform query-similarity estimate. Hence, as SetUni and BagUni it utilizes only inter-document similarities; but, in doing so, BagDupUni rewards to a larger extent documents with multiple instances due to the bag representation and the duplicated instances. The **BagDupMNZ** method integrates also retrieval-scores information by using the retrieval score of a document instance in a new list as the query-similarity estimate of the corresponding node. For $wt^{[1]}$ (i.e., $\lambda = 1$), BagDupMNZ amounts to CombMNZ, as was the case for SetMNZ. (See Appendix A for a proof.) Yet, BagDupMNZ rewards to a larger extent documents with multiple instances than SetMNZ does due to the bag representation of the lists and the duplicated document instances.

## 3. Related Work

Fusion methods often use the ranks of documents in the lists, or their retrieval scores, but not the documents' content (Fox & Shaw, 1994; Voorhees, Gupta, & Johnson-Laird, 1994; Lee, 1997; Vogt & Cottrell, 1999; Croft, 2000b; Dwork et al., 2001; Aslam & Montague, 2001; Montague & Aslam, 2002; Lillis, Toolan, Collier, & Dunnion, 2006; Shokouhi, 2007). For example, Dwork et al. (2001), as us, use Markov chains so as to find prestigious documents in the lists. However, the propagation of relevance status is governed only by information regarding the ranks of documents in the lists. We show in Section 4.2 that using both retrieval scores and inter-document similarities to guide relevance-status propagation is more effective than using each alone. Also, we note that previous work on fusion has demonstrated the relative merits of using retrieval scores rather than rank information (Lee, 1997). Furthermore, as stated in Section 2.2.1, our methods can incorporate and generalize fusion methods that rely on scores/ranks by using the set-based graph representation. We used in Section 2.2.1 the CombSUM and CombMNZ methods, which are based on retrieval scores, as examples. CombSUM is a (non supervised) representative of a general family of linear combination techniques (Vogt & Cottrell, 1999), and CombMNZ is considered a highly effective approach which therefore often serves as a baseline in work on fusion (Lee, 1997; Aslam & Montague, 2001; Montague & Aslam, 2002; Lillis et al., 2006; Shokouhi, 2007). In Section 4.2 we demonstrate the performance merits of our approach with respect to CombSUM and CombMNZ, and additional rank-based fusion methods.

There are several fusion methods that utilize document-based features, some of which are based on the document content, e.g., snippets (summaries) of documents (Lawrence & Giles, 1998; Craswell, Hawking, & Thistlewaite, 1999; Tsikrika & Lalmas, 2001; Beitzel, Jensen, Frieder, Chowdhury, & Pass, 2005; Selvadurai, 2007). However, in contrast to our methods, inter-document similarities were not used in these approaches. Thus, these methods can potentially be incorporated in our fusion framework using the set-based graph representation. Furthermore, we note that our methods can potentially utilize document snippets to estimate inter-document similarities, rather than use the entire document content, if the content is not (quickly) accessible. Indeed, snippets were used for inducing inter-document similarities so as to cluster results of Web search engines (Zamir & Etzioni, 1998).

There is a large body of work on re-ranking an initially retrieved list using graph-based methods that model inter-document similarities within the list (e.g., Daniłowicz & Baliński, 2000; Baliński & Daniłowicz, 2005; Diaz, 2005; Kurland & Lee, 2005, 2006; Zhang et al., 2005; Yang, Ji, Zhou, Nie, & Xiao, 2006). As mentioned in Section 2, our fusion methods could conceptually be viewed as a generalization of some of these approaches (Daniłowicz & Baliński, 2000; Diaz, 2005; Kurland & Lee, 2005); specifically, of methods that utilize both retrieval scores and inter-document-similarities for modeling relevance-status propagation within the list (Daniłowicz & Baliński, 2000; Diaz, 2005). A similar relevance-status propagation method was also employed in work on sentence retrieval for question answering (Otterbacher, Erkan, & Radev, 2005).

Similarities between document headlines were used for merging document lists that were retrieved in response to a query from non-overlapping corpora (Shou & Sanderson, 2002). Specifically, a document is ranked by the sum of similarities between its headline and headlines of other documents. In contrast to our approach, which operates on a single corpus, and which accordingly exploits information regarding multiple occurrences of a document in the lists, retrieval scores were not integrated with these similarities; and, a graph-based approach as we use here was not employed. In Section 4.2 we show that using retrieval scores in the single-corpus-based fusion setup that we explore is highly important; specifically, integrating retrieval scores and inter-document-similarities results in much better performance than that of using only inter-document similarities.

Similarities between documents in (potentially non-overlapping) different corpora were also used to form document *clusters* (Xu & Croft, 1999; Crestani & Wu, 2006) so as to (potentially) improve results browsing (Crestani & Wu, 2006) and to improve collection selection (Xu & Croft, 1999) for search. In contrast to our approach, fusion methods that are based on utilizing information induced from inter-document similarities were not proposed.

There is some recent work on *re-ranking* a retrieved list using inter-document similarities with a second retrieved list (Meister et al., 2010). The idea is that documents that are highly ranked in the original list, and that are similar to documents highly ranked in the second list, should be rewarded. However, in contrast to fusion approaches, documents that are members of the second list, but not of the first list, cannot appear in the final result list. Furthermore, in contrast to our approach, there is no recursive definition for prestige. Most importantly, there is no apparent way of generalizing this method so as to fuse several lists, in contrast to our approach.

Methods utilizing inter-item textual similarities — some using a variant of PageRank as we do here — were also used, for example, for cross-lingual retrieval (Diaz, 2008), prediction of retrieval effectiveness (Diaz, 2007), and text summarization and clustering (Erkan & Radev, 2004; Mihalcea & Tarau, 2004; Erkan, 2006). Specifically, some recent work (Krikon, Kurland, & Bendersky, 2010) has demonstrated the merits of integrating whole-document-based inter-document similarities with inter-passage-similarities for re-ranking a single retrieved list; especially, when using corpora containing long and/or topically heterogeneous documents. Incorporating inter-passage similarities in our methods is a future venue we intend to explore.

## 4. Evaluation

We next study the effectiveness of our similarity-based fusion approach, and the different factors that affect its performance.

### 4.1 Experimental Setup

In what follows we describe the setup used for the evaluation.

#### 4.1.1 Measuring Inter-Document Similarities.

We use a previously proposed language-model-based similarity estimate that was shown to be effective in work on re-ranking a single retrieved list (Kurland & Lee, 2005, 2006, 2010).

Let $p_d^{[\mu]}(\cdot)$ denote the unigram, Dirichlet-smoothed, language model induced from document $d$, where $\mu$ is the smoothing parameter (Zhai & Lafferty, 2001). We set $\mu = 1000$ following previous recommendations (Zhai & Lafferty, 2001). For documents $d_1$ and $d_2$ we define:

$$sim(d_1, d_2) \overset{def}{=} \exp\left(-D\left(p_{d_1}^{[0]}(\cdot) \;\middle\|\; p_{d_2}^{[\mu]}(\cdot)\right)\right);$$

$D$ is the KL divergence. The "closer" the language models of $d_1$ and $d_2$ are, the lower the KL divergence is, and the higher the similarity estimate is.

#### 4.1.2 Data, Evaluation Measures, and Parameters.

We evaluate the performance of our fusion methods using TREC datasets (Voorhees & Harman, 2005), which were also used in some previous work on fusion (e.g., Lee, 1997; Aslam & Montague, 2001; Montague & Aslam, 2002): the ad hoc track of trec3, the web tracks of trec9 and trec10, and the robust track of trec12. Tokenization, Porter stemming, and stopword removal (using the INQUERY list) were applied to documents using the Lemur toolkit[4], which was also used for computing $sim(d_1, d_2)$.

Retrieval methods that utilize inter-document similarities in a query context — e.g., for re-ranking a single retrieved list using graph-based techniques — are known to be most effective when employed over relatively short lists (Willett, 1985; Diaz, 2005; Kurland & Lee, 2010). The reason is that such lists often contain documents that exhibit high surface-level query similarity. Hence, the lists could be thought of as providing "effective" query-based corpus context. Similar arguments were echoed in work on pseudo-feedback-based query expansion (Xu & Croft, 1996; Lavrenko & Croft, 2001; Zhai & Lafferty, 2002; Tao & Zhai, 2006). Furthermore, utilizing inter-document similarities in such short lists was shown to be highly effective in improving precision at the very top ranks (Kurland & Lee, 2005, 2006).[5] Indeed, users of Web search engines, for example, are often interested in the most highly ranked documents (a.k.a., "first page of results"). Given the considerations just mentioned, we take the following design decisions with respect to the evaluation measures that we focus on, the number of lists to be fused, and the number of documents in each list.

---

4. www.lemurproject.org
5. Improving precision at top ranks often results in improving MAP (mean average precision) by the virtue of the way MAP is defined. We show in Section 4.2.1 that our approach improves both precision at top ranks and MAP.

As our focus is on precision at top ranks, we use precision of the top 5 and 10 documents (p@5, p@10) as the main evaluation measures. To determine statistically-significant performance differences, we use the two-tailed Wilcoxon test at the 95% confidence level. This means that, on average, the result of a significance test might be erroneous in one out of every twenty tests. Thus, we employ Bonferroni correction over the corpora per evaluation measure (i.e., a confidence level of 98.75% is also used). Specifically, in the results tables that we present, a statistical-significance mark corresponds to a 95% confidence level; and, the mark is boldfaced if the corresponding performance difference is statistically significant after Bonferroni correction was employed (i.e., using a 98.75% confidence level).

We use our methods to fuse three lists, each of which corresponds to the top-$k$ documents in a submitted run within a track; that is, we use the actual result lists (runs) submitted by TREC's participants. The main focus of our evaluation, up to (and including) Section 4.2.5, is on fusing the three runs that are the most effective among all submitted runs in a track (both automatic and manual); effectiveness is measured by MAP@$k$ , that is, mean average non-interpolated precision at cutoff $k$, henceforth referred to as MAP (Voorhees & Harman, 2005). The three runs to be fused are denoted, by descending order of MAP performance, **run1**, **run2**, and **run3**, respectively. Although MAP is not an evaluation measure we focus on — albeit, we do present MAP performance numbers in Section 4.2.1 — this practice ensures that the initial ranking of the lists to be fused is of relatively high quality; that is, in terms of recall and relative positioning of relevant documents. Yet, the lists to be fused could still be sub-optimal with respect to precision at top ranks. Thus, we use for reference comparisons to our methods the **Optimal Runs** ("opt. run" in short) per evaluation metric and track; that is, for each track, and evaluation metric $m$ (p@5 or p@10), we report the best (average over queries per track) $m$-performance obtained by *any* submitted run in this track. Note that the MAP performance of run1 is the best in a track by the virtue of the way run1 was selected. However, run1 is not necessarily the optimal run with respect to p@5 or p@10. In addition, we compare the performance of our methods with that of the CombSUM and CombMNZ fusion techniques; recall that these methods, which rely solely on retrieval scores, are special cases of some of our methods. In nutshell, we evaluate the effectiveness of our fusion approach, and that of the fusion methods that serve as reference comparisons, in attaining high precision at top ranks with respect to that of (i) the lists to be fused, and (ii) the best performing runs in a track with respect to precision at top ranks.

We note that fusing the three most (MAP) effective runs in a track does not constitute a real-life retrieval scenario as the quality of the lists to be fused is not known in practice, but can rather potentially be predicted (Carmel & Yom-Tov, 2010). Yet, such setup is suitable for a conservative evaluation of our methods, specifically, for studying their ability to effectively fuse lists of high quality. Nevertheless, in Section 4.2.6 we also present the performance of our methods when fusing three runs that are randomly selected from all runs in a track.

Experiments with setting $k$, the number of documents in each list to be fused, to values in $\{10, 20, 30, 40, 50, 75, 100\}$ showed that the fusion methods with which we compare our approach, specifically CombMNZ (Fox & Shaw, 1994; Lee, 1997), often attain (near) optimal precision-at-top-ranks performance for $k = 20$. As it turns out, this is also the case for our most effective fusion methods. Hence, the experiments to follow are based on using the top $k = 20$ documents from each run to be fused. In Section 4.2.4 we present the effect of $k$ on performance.

Our main goal in the evaluation to follow is to focus on the underlying principles of our proposed fusion approach, and on its *potential* effectiveness. We would like to thoroughly compare the different proposed methods of utilizing inter-document similarities, and the factors that affect their performance, rather than engage in excessive performance optimization. With these goals in mind, we start by ameliorating the effects of free-parameter values. We do so by setting the values of the free parameters that our methods incorporate, and those of the reference comparisons, so as to optimize the *average* p@5 performance over the entire set of queries in a track[6]. Thus, we note that the p@10 performance numbers that we present are not necessarily the optimal ones that could be attained. Yet, such an experimental setup is more realistic than that of optimizing performance for each of the evaluation metrics separately. Then, in Sections 4.2.3 and 4.2.4 we present the effect on performance of varying the values of free parameters of our methods. Furthermore, in Section 4.2.5 we present the performance of our most effective methods when values of free parameters are *learned* using cross-validation performed over queries. The value of the ancestry parameter $\alpha$, which is incorporated by our methods, is chosen from $\{5, 10, 20, 30, 40, 50\}$. The value of $\lambda$, which controls the reliance on retrieval scores versus inter-document-similarities, is chosen from $\{0.1, 0.2, \ldots, 1\}$.

For inter-list compatibility of retrieval scores, we normalize the score of a document in a list with respect to the sum of all scores in the list; if a list is of negative retrieval scores, which is usually due to using logs, we use the exponent of a score for normalization[7].

### 4.1.3 Efficiency Considerations.

The number of documents (document instances) in the graphs we construct is at most a few hundreds[8]. Hence, computing inter-document similarities does not incur a significant computational overhead. Even if the entire document content is not quickly accessible, document snippets, for example, could be used for computing inter-document similarities. (This is a future venue we intend to explore.) Similar efficiency considerations were made in work on *clustering* the results retrieved by Web search engines (Zamir & Etzioni, 1998), and in work on re-ranking search results using clusters of top-retrieved documents (Willett, 1985; Liu & Croft, 2004; Kurland & Lee, 2006). In addition, we note that computing prestige over such small graphs takes only a few iterations of the Power method (Golub & Van Loan, 1996).

## 4.2 Experimental Results

We next present the performance numbers of our fusion approach. In Section 4.2.1 we present the main result — the performance of our best-performing models with respect to that of the reference comparisons. Then, in Section 4.2.2 we compare and analyze the performance of all proposed fusion methods. We futher study the merits of using inter-document similarities in Section 4.2.3. The effect on performance of additional factors, e.g., the number of documents in the lists to be fused ($k$), is presented in Section 4.2.4. Section 4.2.5 presents the performance numbers of our most effective models when the values of free parameters are learned using cross validation performed across queries. As noted above, up to (and including) Section 4.2.5, the evaluation is based on fusing the three most effective runs in a track. In Section 4.2.6 we evaluate the performance of our methods when fusing runs that are randomly selected. In Section 4.2.7 we present an analysis of the overlap of relevant and non-relevant documents in the lists to be fused that sheds some more light on the reasons for the relative effectiveness of our approach with respect to that of standard fusion.

---

6. If two parameter settings yield the same p@5, we choose the one *minimizing* p@10 so as to provide conservative estimates of performance.

7. Normalizing retrieval scores with respect to the maximum and minimum scores in a list yields almost exactly the same performance numbers as those we report here.

8. Note that each of the three fused lists contains 20 documents, and each document instance is duplicated, if at all, at most three times.

| | trec3 | | trec9 | | trec10 | | trec12 | |
|---|---|---|---|---|---|---|---|---|
| | p@5 | p@10 | p@5 | p@10 | p@5 | p@10 | p@5 | p@10 |
| opt. run | 76.0 | 72.2 | 60.0 | **53.1** | 63.2 | 58.8 | 54.5 | 48.6 |
| run1 | 74.4 | 72.2 | 60.0 | **53.1** | 63.2 | 58.8 | 51.1 | 44.8 |
| run2 | 72.8 | 67.6 | $45.8^o$ | $38.8^o$ | 54.4 | 50.2 | 52.5 | 48.6 |
| run3 | 76.0 | 71.2 | $38.3^o$ | $34.6^o$ | 55.6 | $46.8^o$ | 51.5 | $45.2^o$ |
| CombSUM | $80.8_{ab}$ | $74.6_{b}$ | $52.9_{bc}$ | $48.5_{bc}$ | $71.2^o_{abc}$ | $\mathbf{61.0}_{bc}$ | 53.7 | $\mathbf{49.2}_{ac}$ |
| BagSum | $\mathbf{83.2}^o_{abc}$ | $78.8^{om}_{abc}$ | $59.6^m_{bc}$ | $48.1_{bc}$ | $71.2^o_{abc}$ | $\mathbf{61.0}_{bc}$ | $55.4_{ac}$ | $\mathbf{49.2}_{ac}$ |
| CombMNZ | $80.8_{ab}$ | $74.6_{b}$ | $55.0_{bc}$ | $48.8_{bc}$ | $71.2^o_{abc}$ | $\mathbf{61.0}_{bc}$ | 53.9 | $\mathbf{49.2}_{ac}$ |
| BagDupMNZ | $\mathbf{83.2}_{ab}$ | $\mathbf{79.0}^{om}_{abc}$ | $\mathbf{60.4}^m_{bc}$ | $47.9_{bc}$ | $\mathbf{72.0}^o_{abc}$ | $\mathbf{61.0}_{bc}$ | $\mathbf{56.6}^m_{abc}$ | $49.0_{ac}$ |

Table 3: Main result table. The performance of two of our most effective fusion methods, BagSum and BagDupMNZ; for $\lambda = 1$, these amount to CombSUM and CombMNZ, respectively. The best performance in each column is boldfaced. Statistically-significant differences with opt. run, run1, run2, and run3, are marked with 'o', a', 'b', and 'c', respectively. (Here and after, we do not mark statistically-significant differences between run1, run2 and run3 to avoid cluttering the presentation, as these convey no additional insight.) Statistically-significant differences between BagSum and CombSUM, and between BagDupMNZ and CombMNZ, are marked with 'm'. The values of $(\lambda, \alpha)$ that yield optimal average p@5 performance for BagSum are $(0.4, 5)$, $(0.6, 40)$, $(0, 5)$ and $(0.1, 5)$ over trec3, trec9, trec10, and trec12, respectively; for BagDupMNZ, $(0.7, 5)$, $(0.1, 20)$, $(0.1, 5)$, and $(0.1, 20)$ yield optimal average p@5 performance for trec3, trec9, trec10, and trec12, respectively.

### 4.2.1 Main Result

Table 3 presents our main result. We present the performance numbers of two of our most effective methods, namely, BagSum and BagDupMNZ. (See Section 4.2.2 for an in-depth analysis of the performance of all our fusion methods.) Recall that for $\lambda = 1$ — i.e., using no inter-document-similarity information — these methods amount to CombSUM and CombMNZ, respectively.

Our first observation based on Table 3 is that in most reference comparisons (track × evaluation measure) the BagSum and BagDupMNZ methods outperform — often to a substantial and statistically-significant degree — each of the three fused runs. Furthermore, in many cases, the performance of our methods is also superior to that of opt. run. This also holds, for example, for p@5 of BagDupMNZ for trec12, a track for which the performance of the runs to be fused (specifically, run2 and run3) can be quite below that of opt. run.

For trec9, our methods' performance is in several cases below that of run1, which is also opt. run with respect to p@5 and p@10. However, these performance differences are not statistically significant. Also, note that run1 is by far more effective than run2 and run3; hence, run2 and run3 potentially have a relatively few relevant documents to "contribute" in addition to those in run1. Nevertheless, the performance of our methods for trec9 is substantially better than that of the other two fused runs (run2 and run3); and, in terms of p@5 — the metric for which performance is optimized — the performance for trec9 of each of BagSum and BagDupMNZ is statistically-significantly better (for BagDupMNZ also after employing Bonferroni correction) than that of its special case, that is, CombSUM and CombMNZ, respectively. The fact that the p@5 performance of CombSUM and CombMNZ is much worse than that of run1 for trec9, which is not the case for other tracks, could be attributed to the fact that the number of relevant documents that are shared among the three runs is the lowest observed with respect to all considered tracks. (We present an analysis of the number of relevant documents shared by the runs in Section 4.2.7.) This is a scenario in which our methods can yield much merit by using inter-document similarities, as is evident in the p@5 performance improvements they post over CombSUM and CombMNZ for trec9.

More generally, we can see in Table 3 that in a majority of the relevant comparisons our methods' performance is superior to that of their special cases that do not utilize inter-document similarities (CombSUM and CombMNZ). The p@5 improvements for trec9, for example, and as noted above, are both substantial and statistically significant. Furthermore, our methods post more statistically significant improvements over the runs to be fused, and over opt. run, than CombSUM and CombMNZ do. Thus, these findings attest to the merits of utilizing inter-document similarities for fusion.

**Analysis of MAP Performance.** Although the focus of the evaluation we present is on precision at top ranks, we are also interested in the general quality of the ranking induced by our methods. Accordingly, we present the MAP performance of the BagSum and BagDupMNZ methods in Table 4. To avoid potential metric-divergence issues (Azzopardi, Girolami, & van Rijsbergen, 2003; Morgan, Greiff, & Henderson, 2004; Metzler & Croft, 2005), that is, optimizing performance for one retrieval metric and presenting performance numbers for a different retrieval metric, we optimize the performance of our methods *only* in this case with respect to MAP.

We can see in Table 4 that except for trec9, our methods outperform — and in quite a few cases, statistically significantly so — the fused runs, and opt. run. (Recall that run1 is the best MAP-performing run in a track; i.e., in terms of MAP, run1 is opt. run.) Moreover, our methods consistently outperform their corresponding special cases, CombSUM and CombMNZ.

**Comparison with Rank-Based Fusion Methods** While the focus of this paper is on fusion methods that utilize retrieval scores, in Table 5 we compare the performance of BagDupMNZ (one of our two best-performing methods) with that of two fusion methods that utilize ranks of documents rather than their scores. The first is a simple round robin approach wherein the order of runs used is run1, run2 and run3. The second rank-based fusion method is Borda (Young, 1974), in which $d$ is scored by the number of documents not ranked higher than it in the lists:

$$P_{Borda}(d) \stackrel{def}{=} \sum_{L_i} \#\{d' \in L_i : S_{L_i}(d') <= S_{L_i}(d)\}.$$

| | trec3 | trec9 | trec10 | trec12 |
|---|---|---|---|---|
| | MAP | MAP | MAP | MAP |
| opt. run | 10.4 | **28.2** | 30.7 | 28.8 |
| run1 | 10.4 | **28.2** | 30.7 | 28.8 |
| run2 | 9.6 | $18.4^{\mathbf{o}}$ | $27.7^{o}$ | 28.4 |
| run3 | 9.5 | $16.8^{\mathbf{o}}$ | $21.6^{\mathbf{o}}$ | 28.1 |
| CombSUM | $10.9_{\mathbf{bc}}$ | $24.9_{\mathbf{bc}}$ | $37.2_{\mathbf{bc}}$ | $30.3^{\mathbf{o}}_{\mathbf{a}}$ |
| BagSum | $11.4^{om}_{\mathbf{abc}}$ | $26.6_{\mathbf{bc}}$ | $37.3_{\mathbf{bc}}$ | $\mathbf{30.5}^{\mathbf{o}}_{\mathbf{a}}$ |
| CombMNZ | $10.9_{\mathbf{bc}}$ | $25.5_{\mathbf{bc}}$ | $37.2_{\mathbf{bc}}$ | $30.3^{\mathbf{o}}_{\mathbf{a}}$ |
| BagDupMNZ | $\mathbf{11.5}^{om}_{\mathbf{abc}}$ | $27.0_{\mathbf{bc}}$ | $\mathbf{38.4}_{\mathbf{bc}}$ | $\mathbf{30.5}^{\mathbf{o}}_{\mathbf{a}}$ |

Table 4: MAP performance numbers. The best performance in each column is boldfaced. Statistically significant differences with opt. run, run1, run2, and run3, are marked with 'o', 'a', 'b', and 'c', respectively. Statistically significant differences between BagSum and CombSUM, and between BagDupMNZ and CombMNZ, are marked with 'm'.

| | trec3 | | trec9 | | trec10 | | trec12 | |
|---|---|---|---|---|---|---|---|---|
| | p@5 | p@10 | p@5 | p@10 | p@5 | p@10 | p@5 | p@10 |
| round robin | 76.4 | 73.2 | 50.4 | 45.6 | 61.6 | 55.2 | 53.9 | 47.3 |
| Borda | 80.4 | 78.6 | 55.0 | **48.3** | 71.2 | **62.0** | 54.3 | 48.8 |
| BagDupMNZ | **83.2** | $\mathbf{79.0}^{\mathbf{r}}_{b}$ | $\mathbf{60.4}^{\mathbf{r}}_{b}$ | 47.9 | $\mathbf{72.0}^{\mathbf{r}}$ | $61.0^{r}$ | **56.6** | **49.0** |

Table 5: Comparison with rank-based fusion methods. Statistically-significant differences between BagDupMNZ and round robin and Borda are marked with 'r' and 'b', respectively.

We can see in Table 5 that BagDupMNZ outperforms both the round robin and the Borda methods in most reference comparisons. Many of the performance differences (especially those with round robin) are quite substantial and some are also statistically significant.

**Upper Bound Analysis**  To study the potential of our approach when completely neutralizing the effects of free-parameter values, we present in Table 6 an upper bound analysis of the p@5 performance of BagDupMNZ. To that end, for each query we use free-parameter values of BagDupMNZ that yield optimized p@5 for this query. Recall that the performance of BagDupMNZ reported above was based on free-parameter values set to optimize average

|  | trec3 | trec9 | trec10 | trec12 |
|---|---|---|---|---|
| OptRunPerQuery | 91.6 | 79.6 | 84.4 | 84.8 |
| opt. run | $76.0^{p}$ | $60.0^{p}$ | $63.2^{p}$ | $54.5^{p}$ |
| run1 | $74.4^{p}$ | $60.0^{p}$ | $63.2^{p}$ | $51.1^{p}$ |
| CombMNZ | $80.8^{p}_{a}$ | $55.0^{p}$ | $71.2^{po}_{a}$ | $53.9^{p}$ |
| BagDupMNZ | $89.6^{o}_{am}$ | $68.3^{po}_{am}$ | $79.6^{o}_{am}$ | $66.1^{po}_{am}$ |

Table 6: Upper bound analysis of the p@5 performance of BagDupMNZ. Specifically, for *each* query we use BagDupMNZ with free-parameter values optimized for p@5 for that query. As a reference comparison, for *each* query we consider the best p@5-performing run (OptRunPerQuery). The performance of run1 (the best (MAP) performing among the three fused runs), opt. run (the run that yields the best average p@5 per track), and CombMNZ is presented for reference. 'p', 'o', 'a', and 'm' mark statistically significant differences with OptRunPerQuery, opt. run, run1, and CombMNZ, respectively.

|  | trec3 | | trec9 | | trec10 | | trec12 | |
|---|---|---|---|---|---|---|---|---|
|  | p@5 | p@10 | p@5 | p@10 | p@5 | p@10 | p@5 | p@10 |
| opt. run | 76.0 | 72.2 | 60.0 | **53.1** | 63.2 | 58.8 | 54.5 | 48.6 |
| run1 | 74.4 | 72.2 | 60.0 | **53.1** | 63.2 | 58.8 | 51.1 | 44.8 |
| run2 | 72.8 | 67.6 | $45.8^{o}$ | $38.8^{o}$ | 54.4 | 50.2 | 52.5 | 48.6 |
| run3 | 76.0 | 71.2 | $38.3^{o}$ | $34.6^{o}$ | 55.6 | $46.8^{o}$ | 51.5 | $45.2^{o}$ |
| SetUni | $79.2_{b}$ | 75.0 | $42.5^{o}_{a}$ | $39.2^{o}_{a}$ | 56.8 | $48.2^{o}_{a}$ | $47.3^{o}$ | $41.5^{o}_{b}$ |
| SetSum | $82.8^{o}_{abc}$ | $78.0^{o}_{abc}$ | $59.2_{bc}$ | $49.2_{bc}$ | $71.2^{o}_{abc}$ | $61.0_{bc}$ | $55.4_{a}$ | $48.5_{ac}$ |
| SetMNZ | $82.0_{ab}$ | $77.2^{o}_{abc}$ | $\mathbf{61.3}_{bc}$ | $49.2_{bc}$ | $71.2^{o}_{abc}$ | $61.0_{bc}$ | $55.6_{ac}$ | $48.5_{ac}$ |
| BagUni | $82.4_{ab}$ | $78.8^{o}_{abc}$ | $59.2_{bc}$ | $47.9_{bc}$ | $70.8_{bc}$ | $\mathbf{61.2}_{bc}$ | 53.1 | 46.5 |
| BagSum | $\mathbf{83.2}^{o}_{abc}$ | $78.8^{o}_{abc}$ | $59.6_{bc}$ | $48.1_{bc}$ | $71.2^{o}_{abc}$ | $61.0_{bc}$ | $55.4_{ac}$ | $\mathbf{49.2}_{ac}$ |
| BagDupUni | $82.0_{ab}$ | $78.6^{o}_{abc}$ | $57.5_{bc}$ | $48.1_{bc}$ | $\mathbf{72.0}^{o}_{abc}$ | $60.4_{bc}$ | 52.9 | 47.8 |
| BagDupMNZ | $\mathbf{83.2}_{ab}$ | $\mathbf{79.0}^{o}_{abc}$ | $60.4_{bc}$ | $47.9_{bc}$ | $\mathbf{72.0}^{o}_{abc}$ | $61.0_{bc}$ | $\mathbf{56.6}_{abc}$ | $49.0_{ac}$ |

Table 7: Performance comparison of all proposed fusion methods. The best result in a column is boldfaced. Statistically significant differences with opt. run, run1, run2, and run3, are marked with 'o', 'a', 'b', and 'c', respectively.

p@5 performance for a track. (The same runs used above are fused here). For reference comparison, we consider for each query the run in the track that yields the best p@5 for that query (denoted OptRunPerQuery). We also present the performance of the opt. run baseline, used above, which is the run that yields the best average p@5 performance per track. The performance of run1 (the most (MAP) effective of the three fused runs) and CombMNZ is presented for reference as well.

As we can see in Table 6 the performance of BagDupMNZ is substantially (and statistically significantly) better than that of opt. run, run1 and CombMNZ with the performance differences being, naturally, much higher than those in Table 3. Thus, we see that using inter-document similarities for fusion can yield substantial merits; and, that optimizing the free-parameter values of our approach per query yields better performance than using the same values for all queries, as could be expected. As already noted, in Section 4.2.5 we study the performance of our approach when using cross-validation to set free-parameter values.

We can also see in Table 6 that except for trec3, the performance of BagDupMNZ is much inferior (and statistically significantly so) to that of selecting the best-performing run per each query (OptRunPerQuery). This is not a surprise as the performance of run1 (the most effective — on average — among the runs to be fused) is also substantially (and statistically significantly) worse than that of OptRunPerQuery; the same observation holds for opt. run, which shows that different runs are the most effective for different queries.

### 4.2.2 Performance Analysis of All Proposed Fusion Methods

In Table 7 we compare the performance of all proposed fusion methods. Our first observation is that using the retrieval scores of documents in the lists, on top of inter-document-similarity information, is important. Indeed, the methods with the suffix "Uni" that use a uniform query-similarity estimate, i.e., that disregard the retrieval scores of documents in the lists, post performance that is almost always worse than that of their counterparts that do utilize retrieval scores for inducing query similarity. (Compare SetUni with SetSum and SetMNZ; BagUni with BagSum; and, BagDupUni with BagDupMNZ.) Furthermore, utilizing retrieval scores results in performance that is almost always better — and in many cases to a statistically significant degree — than that of run2 and run3; the performance also transcends that of run1 and opt. run, except for trec9.

We can also see in Table 7 that the bag representation of the lists yields better performance, in general, than that of using a set representation. (Compare, for example, BagUni with SetUni, and BagSum with SetSum.) Recall that under a bag representation a document is represented by the nodes corresponding to its instances in lists, while under a set representation each document is represented by a single node. Hence, the fact that documents with occurrences in many of the fused lists can draw more prestige-status support via inter-document-similarities than documents with fewer occurrences has positive impact on performance.

Thus, it is not a surprise that the BagSum and BagDupMNZ methods that use a bag-representation of the lists, and which utilize the retrieval scores of documents in the lists, are among the most effective fusion methods that we proposed.

### 4.2.3 The Performance Impact of Using Inter-Document-Similarities

The $\lambda$ parameter in Equation 1 (Section 2) controls the reliance on retrieval scores versus inter-document similarity information. Setting $\lambda = 1$, i.e., using no inter-document-similarity information, results in the "XSum" methods being equivalent to CombSUM, and the "XMNZ" methods being equivalent to CombMNZ. In Table 3 we showed that BagSum outperforms CombSUM and that BagDupMNZ outperforms CombMNZ. We now turn to study the performance of all "XSum" and "XMNZ" methods with respect to their "special cases", that is, CombSUM and CombMNZ, respectively.

We can see in Table 8 that in a majority of the relevant comparisons (track × evaluation metric), each of our methods outperforms its special case, with several of the differences being statistically significant. We therefore conclude that inter-document-similarities are indeed a helpful source of information for fusion.

| | trec3 | | trec9 | | trec10 | | trec12 | |
|---|---|---|---|---|---|---|---|---|
| | p@5 | p@10 | p@5 | p@10 | p@5 | p@10 | p@5 | p@10 |
| CombSUM | 80.8 | 74.6 | 52.9 | 48.5 | 71.2 | **61.0** | 53.7 | **49.2** |
| SetSum | 82.8 | $78.0^m$ | $59.2^m$ | **49.2** | 71.2 | **61.0** | 55.4 | 48.5 |
| BagSum | **83.2** | **78.8$^{\mathbf{m}}$** | $59.6^m$ | 48.1 | 71.2 | **61.0** | 55.4 | **49.2** |
| CombMNZ | 80.8 | 74.6 | 55.0 | 48.8 | 71.2 | **61.0** | 53.9 | **49.2** |
| SetMNZ | 82.0 | 77.2 | **61.3$^m$** | **49.2** | 71.2 | **61.0** | 55.6 | 48.5 |
| BagDupMNZ | **83.2** | **79.0$^{\mathbf{m}}$** | $60.4^m$ | 47.9 | **72.0** | 61.0 | **56.6$^{\mathbf{m}}$** | 49.0 |

Table 8: Comparison of the similarity-based fusion methods with their special cases, Comb-SUM and CombMNZ. Best performance in a column is boldfaced. Statistically significant difference between a method and its special case is marked with 'm'.

We further study the effect of varying the value of $\lambda$ on the p@5 performance of one of our two most effective methods, BagDupMNZ, in Figure 1. Our first observation is that except for trec9, and for all values of $\lambda$, BagDupMNZ yields performance that transcends that of run1, which is the most effective among the three fused runs; for most values of $\lambda$ the performance of BagDupMNZ is also better than that of opt. run. Trec9 is an exception in that BagDupMNZ outperforms run1, which is also opt. run, for a single value of $\lambda$. Recall that for trec9 the performance of run1 is by far better than that of the other fused runs.

Another observation that we make based on Figure 1 is that for most tracks $\lambda \geq 0.6$ yields better performance than that attained by using lower values of $\lambda$. This finding further demonstrates the importance of utilizing retrieval scores of documents as specified above. For $\lambda = 1$ no inter-document-similarities are used and BagDupMNZ amounts to CombMNZ. We can also see that in many cases wherein $\lambda \in \{0.7, 0.8, 0.9\}$ BagDupMNZ outperforms CombMNZ; for trec9 and trec12 these improvements are quite substantial. These findings echo those specified above with regard to the merits of utilizing inter-document-similarities for fusion. Finally, we note that the performance merits attained by using inter-document similarities are even more emphasized when the runs to be fused are randomly selected from all those available for a track (as we will show in Section 4.2.6), rather than being the best (MAP) performing ones as those used here.

### 4.2.4 FURTHER ANALYSIS

**Effect of $\alpha$.** The similarity-based fusion methods incorporate two free parameters: $\lambda$, which controls the reliance on retrieval scores versus inter-document-similarities; the effect of $\lambda$ was studied above; and, $\alpha$, which is the number of nearest neighbors considered for a node in the graphs we use. In Figure 2 we analyze the effect of $\alpha$ on the p@5 performance of BagDupMNZ.

We can see in Figure 2 that small values of $\alpha$ ($\in \{5, 10, 20\}$) often yield better performance than larger values. The same finding was reported in work on utilizing nearest-neighbors-based graphs for re-ranking a single retrieved list (Diaz, 2005; Kurland & Lee, 2005). Furthermore, we can see that small values of $\alpha$ yield performance that transcends that of run1 and opt. run, except for trec9. Another observation that we make based on Figure 2 is that for most corpora, and most values of $\alpha$, BagDupMNZ outperforms its special case, CombMNZ.

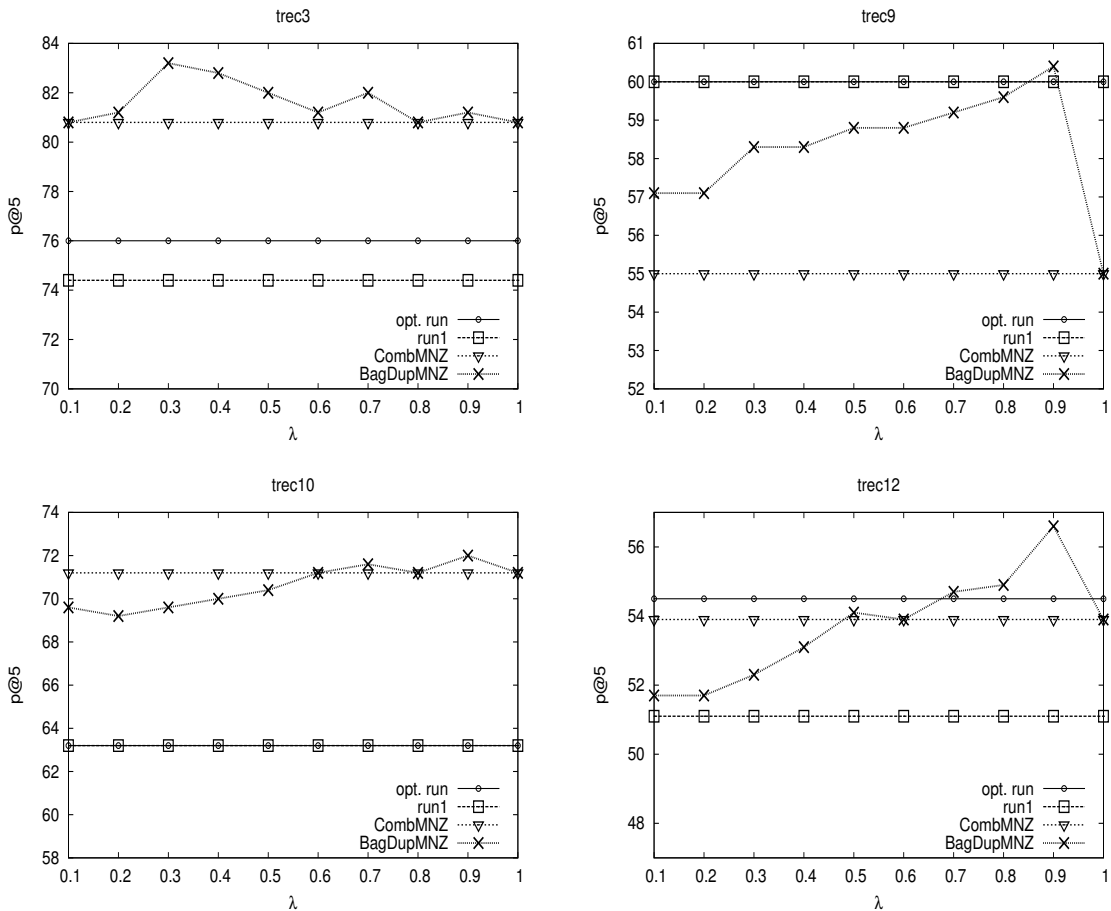

Figure 1: Effect of varying the value of $\lambda$ (refer to Equation 1 in Section 2) on the p@5 performance of BagDupMNZ; $\lambda = 1$ amounts to CombMNZ. The performance of opt. run, run1 and CombMNZ is depicted with horizontal lines for reference. Note: figures are not to the same scale.

**Effect of $k$.** The experimental design that was used insofar, and which was presented in Section 4.1, was based on the observation that attaining high precision at top ranks calls for fusion of relatively short retrieved lists. Indeed, the performance numbers presented above demonstrated the effectiveness of fusing lists of 20 documents each. In Figure 3 we present the effect of $k$ (the number of documents in each retrieved list) on the p@5 performance of BagDupMNZ and CombMNZ.

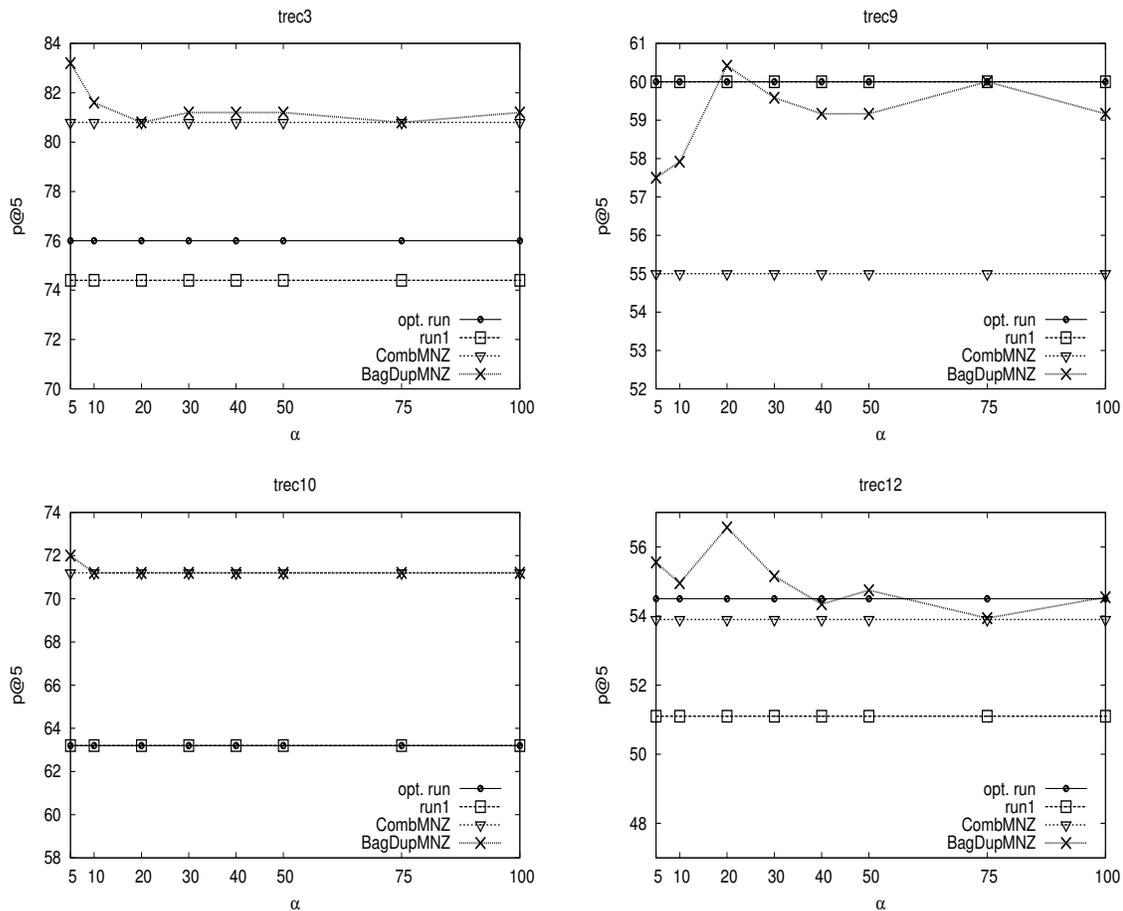

Figure 2: Effect of $\alpha$ on p@5 performance. The performance of opt. run, run1 and CombMNZ is depicted with horizontal lines for reference. Note: figures are not to the same scale

We can see in Figure 3 that for almost all values of $k$ the performance of BagDupMNZ transcends that of CombMNZ. This finding also holds with respect to opt. run, except for the trec9 case. These findings further attest to the merits of utilizing inter-document similarities for fusion. Furthermore, small values of $k$, specifically $k = 20$ which was used heretofore, often yield (near) optimal performance for both BagDupMNZ and CombMNZ. Thus, we indeed see that fusing short lists, specifically, when utilizing inter-document similarities, often leads to very effective precision-at-top-ranks performance.

### 4.2.5 Learning Free-Parameter Values

The performance numbers presented insofar were based on free-parameter values that yield optimal average p@5 performance with respect to the set of queries for track. This experimental setup enabled us to study the potential performance of our approach, and to carefully analyze the different factors that affect it.

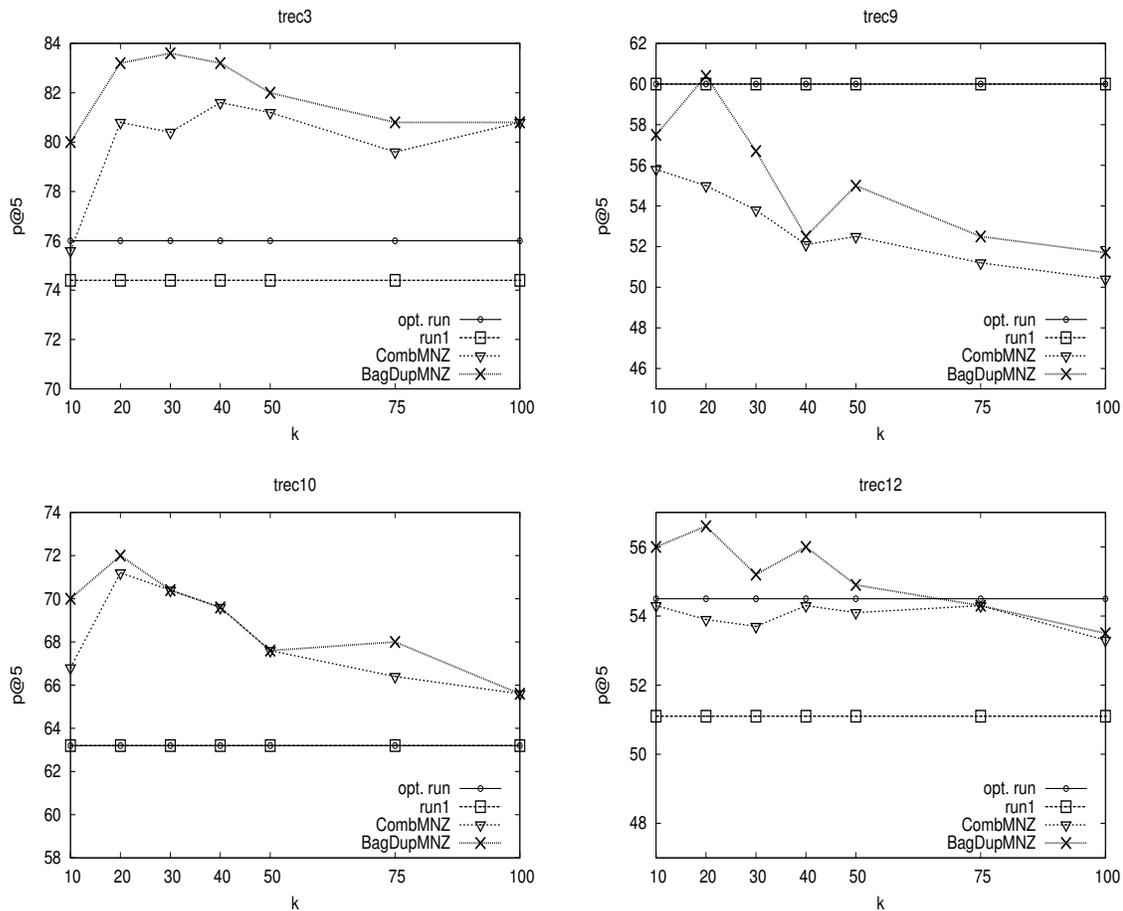

Figure 3: Effect of varying $k$, the number of documents in each fused list (run), on the p@5 performance of BagDupMNZ. The performance of opt. run, run1, and CombMNZ is depicted for reference. Note: figures are not to the same scale.

Now, we turn to explore the question of whether effective values of the free parameters of our methods, $\lambda$ and $\alpha$, generalize across queries; that is, whether such values can be learned[9]. To that end, we employ a leave-one-out cross validation procedure wherein the free parameters of a method are set for each query to values that optimize average p@5 performance over all other queries in a track. The resultant performance numbers of our best performing methods, BagSum and BagDupMNZ, are presented in Table 9.

We can see in Table 9 that both BagSum and BagDupMNZ post better performance, in a vast majority of the relevant comparisons (track × evaluation measure), than that of opt. run, and that of the three runs that are fused; many of these performance improvements are statistically significant.

---

9. Note that such analysis is different than that of studying the effect of free-parameter values on the average performance that was presented above.

| | trec3 | | trec9 | | trec10 | | trec12 | |
|---|---|---|---|---|---|---|---|---|
| | p@5 | p@10 | p@5 | p@10 | p@5 | p@10 | p@5 | p@10 |
| opt. run | 76.0 | 72.2 | 60.0 | **53.1** | 63.2 | 58.8 | 54.5 | 48.6 |
| run1 | 74.4 | 72.2 | 60.0 | **53.1** | 63.2 | 58.8 | 51.1 | 44.8 |
| run2 | 72.8 | 67.6 | $45.8^o$ | $38.8^o$ | 54.4 | 50.2 | 52.5 | 48.6 |
| run3 | 76.0 | 71.2 | $38.3^o$ | $34.6^o$ | 55.6 | $46.8^o$ | 51.5 | $45.2^o$ |
| CombSUM | $80.8_{ab}$ | $74.6_b$ | $52.9_{bc}$ | $48.5_{bc}$ | $\mathbf{71.2}^o_{abc}$ | $61.0_{bc}$ | 53.7 | $\mathbf{49.2_{ac}}$ |
| BagSum | $\mathbf{83.2}^o_{abc}$ | $78.8^{om}_{abc}$ | $57.9_{bc}$ | $47.3_{bc}$ | $67.6^m_{bc}$ | $\mathbf{61.4_{bc}}$ | $54.7_a$ | $\mathbf{49.2_{ac}}$ |
| CombMNZ | $80.8_{ab}$ | $74.6_b$ | $55.0_{bc}$ | $48.8_{bc}$ | $\mathbf{71.2}^o_{abc}$ | $61.0_{bc}$ | 53.9 | $\mathbf{49.2_{ac}}$ |
| BagDupMNZ | $82.4_{ab}$ | $79.0^{om}_{abc}$ | $\mathbf{60.4}^m_{bc}$ | $47.9_{bc}$ | $70.8_{bc}$ | $60.4_{bc}$ | $\mathbf{56.6}^m_{abc}$ | $49.0_{ac}$ |

Table 9: Performance numbers when employing leave-one-out cross validation to set free-parameter values. The best performance in each column is boldfaced. Statistically-significant differences with opt. run, run1, run2, and run3, are marked with 'o', 'a', 'b', and 'c', respectively. Statistically significant differences between BagSum and CombSUM, and between BagDupMNZ and CombMNZ, are marked with 'm'.

We next compare our methods with their special cases that do not utilize inter-document similarities. That is, we compare BagSum with CombSUM and BagDupMNZ with CombMNZ. With respect to p@5 — the metric for which performance was optimized in the learning phase — our methods outperform their special cases for all tracks, except for that of trec10; some of these improvements are also statistically significant (e.g., refer to BagDupMNZ versus CombMNZ on trec9 and trec12). Furthermore, we note that most cases in which CombSUM and CombMNZ outperform our methods are for p@10. We attribute this finding to the metric divergence issue (Azzopardi et al., 2003; Morgan et al., 2004; Metzler & Croft, 2005) — optimizing performance in the learning phase with respect to one metric (p@5 in our case), and testing the performance with respect to another metric (p@10 in our case), albeit somewhat connected. Recall that while our methods incorporate two free parameters, the CombSUM and CombMNZ methods do not incorporate free parameters. Additional examination of Table 9 reveals that our methods post more statistically significant improvements over the runs to be fused and opt. run than CombSUM and CombMNZ do.

All in all, these results demonstrate the effectiveness of our methods when employing cross validation so as to set free-parameter values.

### 4.2.6 Fusing Randomly Selected Runs

Heretofore, the evaluation of our approach was based on fusing the most (MAP) effective runs in a track. We now turn to study the effectiveness of our best performing fusion methods when fusing randomly selected runs.

We select 20 random triplets of runs from each track. The best performing run among the three is denoted run1, the second best is denoted run2, and the worst among the three is denoted run3. We then fuse the three runs using either the standard fusion methods, CombSUM and CombMNZ, or our methods that generalize these, namely, BagSum and BagDupMNZ, respectively. The performance numbers presented in Table 10 represent averages over the 20 samples.[10] The free parameters of BagSum and BagDupMNZ were set to values optimizing average p@5 performance over queries for a track per each triplet of runs. (The optimization procedure described in Section 4.1 was used.) Statistically significant differences between two methods are determined based on the average (over 20 samples) performance per each query.

| | trec3 | | trec9 | | trec10 | | trec12 | |
|---|---|---|---|---|---|---|---|---|
| | p@5 | p@10 | p@5 | p@10 | p@5 | p@10 | p@5 | p@10 |
| run1 | 68.9 | 57.4 | **22.1** | **19.6** | 32.7 | 28.5 | 46.0 | 39.9 |
| run2 | 57.4 | 55.4 | 16.2 | 14.7 | 28.5 | 24.9 | 39.9 | 34.4 |
| run3 | 42.3 | 41.4 | 10.9 | 10.2 | 18.3 | 16.0 | 27.4 | 23.2 |
| CombSUM | $65.6_{bc}$ | $61.3_{abc}$ | $19.6_{abc}$ | $17.6_{abc}$ | $32.4_{bc}$ | $28.3_{bc}$ | $44.4_{abc}$ | $37.7_{abc}$ |
| BagSum | $\mathbf{76.1}^{m}_{bc}$ | $\mathbf{70.4}^{m}_{abc}$ | $22.0^{m}_{abc}$ | $18.2^{m}_{abc}$ | $\mathbf{36.6}^{m}_{abc}$ | $\mathbf{30.5}^{m}_{abc}$ | $\mathbf{47.8}_{bc}$ | $\mathbf{40.6}^{m}_{bc}$ |
| CombMNZ | $65.7_{bc}$ | $61.3_{abc}$ | $20.0_{abc}$ | $17.5_{abc}$ | $33.6_{bc}$ | $28.7_{bc}$ | $44.4_{abc}$ | $37.7_{abc}$ |
| BagDupMNZ | $75.7^{m}_{bc}$ | $70.1_{bc}$ | $21.3^{m}_{abc}$ | $18.0^{m}_{abc}$ | $36.5^{m}_{abc}$ | $30.2^{m}_{abc}$ | $46.6_{bc}$ | $40.4_{bc}$ |

Table 10: Fusing randomly selected runs. The performance numbers represent averages over 20 random samples of triplets of runs. The best performance in a column is boldfaced. Statistically significant differences of a fusion method with run1, run2, and run3, are marked with 'a', 'b', and 'c', respectively. Statistically significant differences between BagSum and CombSUM, and between BagDupMNZ and CombMNZ, are marked with 'm'.

Our first observation based on Table 10 is that our methods are highly effective in fusing randomly selected runs. In almost all reference comparisons (track × evaluation measure), they outperform each of the three fused runs; most of these improvements are substantial and statistically significant. The only exception is for run1 for trec9, which outperforms all fusion methods.

We can also see in Table 10 that BagSum is slightly more effective than BagDupMNZ. However, when fusing the best-performing runs in a track, as was the case above, the picture was somewhat reversed. We attribute this finding to the relatively low overlap of relevant documents in the randomly selected runs. Specifically, we show below that this overlap is much smaller than that for the best-performing runs. Thus, the use of information regarding multiple appearances in lists, which is quite emphasized by BagDupMNZ, is not of significant merit. Note that this also holds for the standard fusion methods. That is, the superiority of CombMNZ to CombSUM — the former emphasizes appearances in multiple lists more than the latter — is less substantial than that for fusing the best performing runs.

Perhaps the most important observation that we can make based on Table 10 is that our methods are always more effective than the standard fusion approaches, which constitute their special cases; that is, compare BagSum with CombSUM and BagDupMNZ with CombMNZ. Many of the performance differences are also statistically significant. Furthermore, in most relevant comparisons, CombSUM and CombMNZ are outperformed by run1 — the best performing run among the three fused — while the reverse holds for our methods. Thus, these results support the merits of utilizing inter-document similarities for fusion.

---

10. We note that the drop in performance when moving from run1 to run2 and run3 is the highest for trec9. This is because many runs in trec9 are of very low quality and contain very few relevant documents if any (Meister et al., 2010).

| | trec3 | | | | | | trec9 | | | | | |
|---|---|---|---|---|---|---|---|---|---|---|---|---|
| | Rel | | | Non-Rel | | | Rel | | | Non-Rel | | |
| | 1 | 2 | 3 | 1 | 2 | 3 | 1 | 2 | 3 | 1 | 2 | 3 |
| Best runs | 59.2 | 24.6 | 16.2 | 81.1 | 14.5 | 4.3 | 61.4 | 25.3 | 13.3 | 79.4 | 14.0 | 6.6 |
| Random runs | 66.9 | 25.3 | 7.7 | 84.9 | 12.9 | 2.2 | 78.6 | 24.4 | 6.3 | 78.6 | 17.8 | 3.6 |

| | trec10 | | | | | | trec12 | | | | | |
|---|---|---|---|---|---|---|---|---|---|---|---|---|
| | Rel | | | Non-Rel | | | Rel | | | Non-Rel | | |
| | 1 | 2 | 3 | 1 | 2 | 3 | 1 | 2 | 3 | 1 | 2 | 3 |
| Best runs | 56.9 | 26.6 | 16.5 | 77.4 | 16.0 | 6.6 | 32.6 | 24.6 | 42.8 | 51.9 | 23.3 | 24.8 |
| Random runs | 66.6 | 22.8 | 10.6 | 79.6 | 14.9 | 5.5 | 48.5 | 27.8 | 23.6 | 68.0 | 20.0 | 12.4 |

Table 11: The percentage of (non-) relevant documents (of those that appear in at least one of the three runs to be fused) that appear in one (1), two (2) , or all three (3) runs. The number of documents, $k$, considered for each run is 20. The three runs are either the best (MAP) performing in the track, or randomly selected; in the latter case, percentages represent averages over 20 random samples. Percentages may not sum to 100 due to rounding.

### 4.2.7 Analysis of (non-) Relevant Documents Overlap in the Lists

The results just presented show that using inter-document-similarities is highly effective when fusing randomly selected runs. In fact, the relative performance improvements over the standard fusion methods that do not utilize inter-document similarities are larger than those observed above when fusing the best-performing runs. Furthermore, the findings presented above attested to the relative limited merit of heavily emphasizing information regarding multiple appearances of documents in the randomly selected runs with respect to the case with the best-performing runs. Hence, we turn to analyze the relevant and non-relevant document overlap between runs when using randomly-selected runs and when using the best-performing runs.

In Table 11 we present the percentage of (non-) relevant documents, of those appearing in at least one of the three runs to be fused, that appear in one, two, or all three runs. We use the top-20 documents from each run as above. We present percentages for the three best-performing runs in a track and for three randomly selected runs; in the latter case, we report averages over 20 samples of triplets of runs. In all cases, percentages are averages over all queries in a track.

Our first observation based on Table 11 is that for most tracks, a majority of the relevant documents appears in only one of the three runs. This finding supports the motivation for our approach; that is, using inter-document similarities so as to transfer relevance-status support between different (similar) relevant documents across the lists. We can also see that the same finding holds for non-relevant documents — a majority of non-relevant documents appears in only one of the three fused runs. We note that previous reports, supporting to a certain extent the cluster hypothesis, have already shown that a majority of the nearest neighbors of a relevant document in the similarity space tend to be relevant; while, those of a non-relevant document tend to be both relevant and non-relevant (Kurland, 2006). This study was performed upon documents retrieved in response to a query as is the case here. Hence, while relevant documents tend to maintain most prestige-status support within the set of relevant documents, non-relevant documents tend to spread this support among relevant and non-relevant documents. Furthermore, the percentage of non-relevant documents that appears in exactly one run is larger than that for relevant documents. This finding echoes those used to explain the effectiveness of standard fusion methods that emphasize appearance in many lists — i.e., that the overlap between relevant documents in the lists is higher than that for non-relevant documents (Lee, 1997).

We can also see in Table 11 that the percentage of relevant documents that appear in only one run is much larger when using randomly selected runs than when using the best-performing runs. In other words, the relevant-document overlap across lists in the best-performing-runs case is higher than that in the randomly-selected runs case. This finding helps to explain the observations made above: (i) the relative performance gains posted by our methods with respect to the standard fusion approaches, which do not utilize inter-document similarities, are larger for randomly selected runs than for the best performing runs, and (ii) heavily emphasizing document appearances in multiple runs is not as effective in the random-runs case as it is in the best-runs case.

To further explore the findings just stated, we present in Figure 4 the percentage of relevant documents that appear in only one of the three runs as a function of the number of documents ($k$) in each run. It is evident in Figure 4 that for the best-performing runs case the percentages are lower than for the randomly-selected runs case, for most values of $k$. This finding further supports the conclusion above with regard to the relative effectiveness of our approach with the best-performing runs versus randomly-selected runs. Furthermore, for most tracks, and for most values of $k$, at least 40% of the relevant documents in the runs to be fused appear in only one of the three runs. This finding further demonstrates the mismatch between relevant-document sets in the runs — a scenario motivating the development of our fusion approach.

## 5. Conclusion and Future Work

We presented a novel approach to fusing document lists that were retrieved in response to a query. Our approach lets similar documents across (and within) lists provide relevance status support to each other. We use a graph-based method to model the propagation of relevance status between documents in the lists. The propagation is governed by inter-document-similarities and by the retrieval scores of documents in the lists.

Empirical evaluation demonstrated the effectiveness of our approach. We showed that our methods are highly effective in fusing TREC runs. This finding holds whether the runs are the most effective per TREC's track (challenge), or randomly selected from the track. We also showed that the performance of our methods transcends that of effective standard fusion methods that utilize only retrieval scores or ranks of documents.

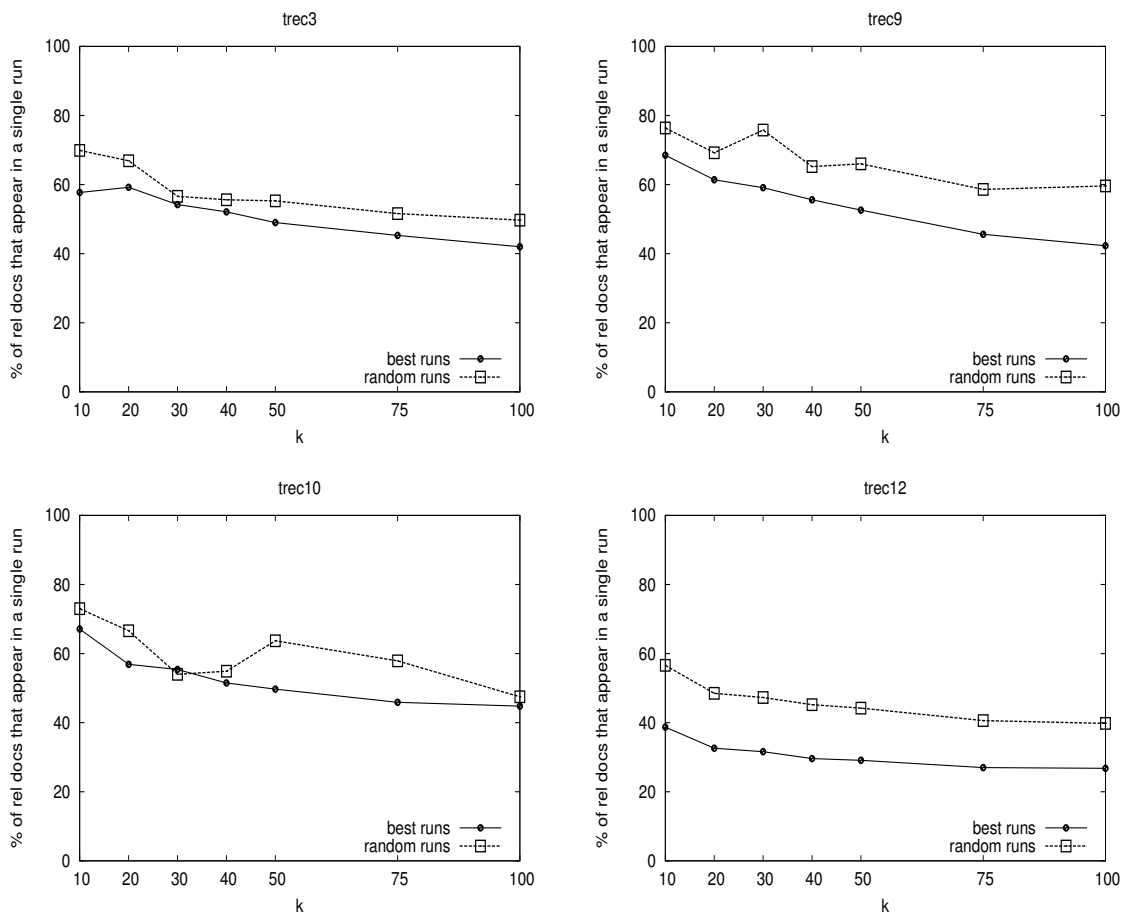

Figure 4: The percentage of relevant documents (of those that appear in at least one of the three runs to be fused) that appear in only one of the runs as a function of the number of documents in a run ($k$). The runs are either the best-performing in a track, or randomly selected; in the latter case, numbers represent averages over 20 random samples.

One family of our proposed methods, namely, the set-based family, can incorporate any fusion method that relies on retrieval scores/ranks. More specifically, we showed that if inter-document-similarities are not utilized, then these set-based methods reduce to the standard fusion method that they incorporate. We have used the CombSUM and CombMNZ fusion methods as examples for instantiating set-based fusion approaches. Naturally then, utilizing additional fusion methods that rely on retrieval scores/ranks is a future venue we intend to explore.

Another venue we intend to explore is the effect of our approach on the diversity of results in the final result list (Carbonell & Goldstein, 1998); and, exploring ways to adapt our methods so as to improve *aspect coverage* in the result list (Zhai, Cohen, & Lafferty, 2003).

**Acknowledgments**

We thank the reviewers for their helpful comments. We also thank Malka Gorfine for her helpful comments. This paper is based upon work supported in part by the Israel Science Foundation under grant no. 557/09, and by IBM's SUR award. Any opinions, findings and conclusions or recommendations expressed in this material are the authors' and do not necessarily reflect those of the sponsoring institutions.

## Appendix A

**Proposition 1.** *Let $f$ be some fusion method that is based on retrieval scores/ranks, e.g., CombSUM or CombMNZ; $f(d)$ is the score assigned by $f$ to document $d$ that appears in at least one of the lists to be fused. Suppose we use $f(d)$ as the query-similarity estimate of $d$ in the set-based group of methods, that is, $\widehat{sim}(d,q) \stackrel{def}{=} f(d)$. Then, using $wt^{[1]}$ (i.e., setting $\lambda = 1$ in Equation 1) results in the final retrieval score of $d$ in Table 1 (Score($d$)) being rank-equivalent to $f(d)$.*

*Proof.* Each node $v$ in the graph corresponds to a *different* document $d$, and has $|V|$ incoming edges the weight of each is $\frac{\widehat{sim}(d,q)}{\sum_{v' \in V} \widehat{sim}(v',q)}$; $\widehat{sim}(d,q)$ is $f(d)$. Hence, this weight is the unique solution to Equation 2, which serves as $d$'s final retrieval score, and is rank-equivalent to $f(d)$.

**Proposition 2.** *Using $wt^{[1]}$ in the BagSum algorithm amounts to the CombSUM algorithm.*

*Proof.* A node $v$ in the graph corresponds to a document-instance $L_i^j$ of some document $d$; and, has $|V|$ incoming edges, the weight of each is $\frac{S(L_i^j)}{\sum_{v' \in V} \widehat{sim}(v',q)}$. This weight is, therefore, the prestige score $P(L_i^j; G^{[\lambda]})$ of $v$ as computed in Equation 2 . By definition, the final retrieval score of $d$ is $\sum_{i,j:L_i^j \equiv d} P(L_i^j; G^{[\lambda]})$. This score is (following the definitions and the above) $\sum_{L_i:d \in L_i} \frac{S_{L_i}(d)}{\sum_{v' \in V} \widehat{sim}(v',q)}$, which is rank-equivalent to $P_{CombSUM}(d)$.

**Proposition 3.** *Using $wt^{[1]}$ in the BagDupMNZ algorithm amounts to the CombMNZ algorithm.*

*Proof.* Let $L_i^j$ be a document-instance of document $d$. Following the definitions and Proposition 2, the prestige value of a single copy of $L_i^j$ from the newly defined list is $\frac{S_{L_i}(d)}{\sum_{v' \in V} \widehat{sim}(v',q)}$. There are $n = \#\{L_i : d \in L_i\}$ such copies in the new list. Therefore, by definition, the final retrieval score of $d$ is $n \sum_{L_i:d \in L_i} \frac{S_{L_i}(d)}{\sum_{v' \in V} \widehat{sim}(v',q)}$, which is rank equivalent to $P_{CombMNZ}(d)$.